# High Sensitivity Snapshot Spectrometer Based on Deep Network Unmixing

Xiaoyu Chen,[1,co] Xu Wang,[1,co] Lianfa Bai, Jing Han,[*] and Zhuang Zhao[*]

*School of Electronic Engineering and Optoelectronic Technology, Nanjing University of Science and Technology, Nanjing 210094, China*

*Corresponding author: eohj@njust.edu.cn, Zhaozhuang@njust.edu.cn*

[1.] These authors contributed equally to this work.

**Abstract:** In this paper, we present a convolution neural network based method to recover the light intensity distribution from the overlapped dispersive spectra instead of adding an extra light path to capture it directly for the first time. Then, we construct a single-path sub-Hadamard snapshot spectrometer based on our previous dual-path snapshot spectrometer. In the proposed single-path spectrometer, we use the reconstructed light intensity as the original light intensity and recover high signal-to-noise ratio spectra successfully. Compared with dual-path snapshot spectrometer, the network based single-path spectrometer has a more compact structure and maintains snapshot and high sensitivity. Abundant simulated and experimental results have demonstrated that the proposed method can obtain a better reconstructed signal-to-noise ratio spectrum than the dual-path sub-Hadamard spectrometer because of its higher light throughput.

## 1. Introduction

Spectral detection is widely used in industry, remote sensing, military, and many other aspects. However, the traditional slit spectrometer has a contradiction between fast acquisition and high signal-to-noise ratio (SNR). If we want to improve the response speed of the system, it will inevitably reduce the acquisition time of the system, resulting in a decrease in the system SNR. In order to improve the SNR of spectral detection, researchers have introduced multiplexing technology, but traditional multiplexing technology requires multiple measurements to obtain the final results, which cannot be used in snapshot measurement scenarios. Therefore, this paper uses the multiplexed snapshot spectroscopy measurement technology based on our previous work [1]. This method needs to measure both the light intensity distribution and the overlapped dispersive spectra. The spectrum to be measured can be obtained by reversible reconstruction, and finally, the spectral data with a higher SNR than the traditional spectrometer can be obtained. However, there are some problems in the practical application of this method. Firstly, the simultaneous acquisition of light intensity and dispersion image requires a dual optical path system, which will reduce the overall light throughput by half and decrease the SNR. Secondly, dual-camera images require pixel-level registration, it will cause a lot of trouble and affecting the reconstruction results. Finally, the dual optical path system increases the overall system size and reduces reliability. Therefore, how to use a single camera to complete the reconstruction process while ensuring the quality of the reconstructed spectrum becomes a problem must be solved.

In order to complete the reconstruction process with a single camera, the original intensity distribution must be reconstructed from the captured overlapped dispersive spectra. At present, the methods of obtaining the original light intensity from spectral images mainly including traditional spectral imaging, compressed sensing spectral imaging and so on. However, there are scanning or multiple encoding processes in traditional spectral imaging, which are not suitable for snapshot spectral acquisition.

Compressed sensing is a representative method to realize snapshot acquisition. Compressed sensing was first proposed by Donoho, Candes, and Tao in 2006[2,3]. In the same year, Brady et al. of Duke University proposed a new spectral imaging technology called Coded Aperture Snapshot Spectral Imager (CASSI) and developed several CASSI-based snapshot imaging spectrometers in the following years [4-6]. In the aspect of reconstruction algorithms, greedy algorithm [7-9], $L_1$ norm convex optimization algorithm [10-12], $L_1$ norm non-convex optimization algorithm [13], Bayesian method [14-16] and other reconstruction algorithms have been developed. However, these reconstruction algorithms all have some drawbacks. Although the greedy algorithm has low computational complexity, the reconstruction effect is not ideal. The $L_1$ norm minimization method has good reconstruction performance, but it has high computational complexity. The Bayesian method lacks a strict theoretical guarantee. Thus, we should develop new reconstruction algorithms.

Based on the idea of compressed sensing, researchers proposed a static multimode multichannel spectrometer (MMS), which takes advantage of the noise reduction performance of Hadamard coding and uses the non-negative least squares method to directly reconstruct the spectrum [17-19]. This method can achieve fast measurement and maintain the advantage of high light throughput. However, compressed sensing is an ill-posed inverse problem, and the measurement results are not stable enough, which limits its application.

In our previous work, we proposed a scheme of snapshot HTS spectral measurement based on dual optical path design. In this snapshot spectrometer, all incident light is encoded simultaneously by the encoding matrix, and a special optical path is designed to measure the intensity distribution of the scene. The spectral measurement problem is transformed into a positive definite problem so that a stable and reliable SNR enhancement [20,21] can be obtained.

As mentioned above, the light intensity on the spectral detection path will be reduced by half. At the same time, the light intensity measurement path needs pixel-level registration, and the overall structure is complex. In recent years, the convolution neural network has shown strong fitting ability in many fields. In order to further enhance the intensity of the spectral detection light path and reduce the complexity of the system, we designed a convolution neural network to restore the light intensity distribution through the overlapped dispersive spectra and realize the single-path high-sensitivity snapshot spectral measurement.

The main contributions of this paper are as follows:

(1) Design a neural network to solve the problem of spectral image unmixing and realize the direct acquisition of light intensity distribution through an overlapped dispersive spectral image.

(2) Prove the feasibility of using reconstructed light intensity data for sub-Hadamard matrix spectral detection.

(3) Both simulated and experimental results demonstrated the performance of sub-Hadamard matrix snapshot spectrum detection based on network reconstruction can achieve similar or even better reconstructed results compared with the dual-path scheme and improving the availability of the system on the premise of reducing the extra optical path.

## 2. Design of snapshot spectrum detection framework

We take 7×7 Hadamard matrix as an example to explain the spectral dispersion overlapping process in the snapshot spectrometer and show it in the following figure.

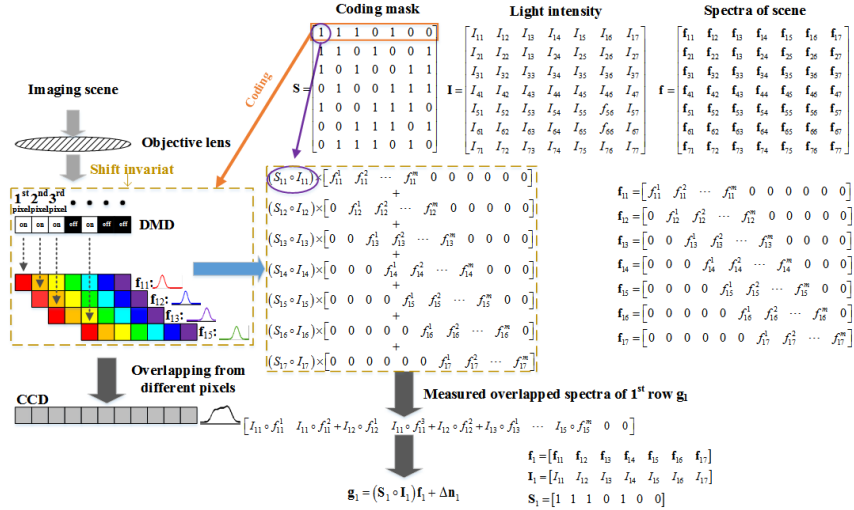

Figure 1. Principle of Hadamard transform spectrum measurement

In Figure 1, $S_1$ is $1^{st}$ row of Hadamard-S matrix, $I_1$ is the $1^{st}$ row light intensity distribution of the scene, $f_1$ is the spectrum of $1^{st}$ row, $f_{11}$ is the spectrum of $1^{st}$ pixel in $1^{st}$ row, $m$ represents its range, and the zeros in the spectrum represents shift invariant, which makes processing easy to understand. The measurement problems of $i^{th}$ row of snapshot HTS are as follows:

$$\mathbf{g}_i = (\mathbf{S}_i \circ \mathbf{I}_i)\mathbf{f}_i + \mathbf{n}_{\text{snap}_i} \qquad (1)$$

Where, $\mathbf{g}_i$ is the overlapping dispersion spectrum of the $i^{th}$ row, $\mathbf{S}_i$ is the $i^{th}$ row of Hadamard-S matrix, $\mathbf{I}_i$ is the $i^{th}$ row of light intensity distribution, $\mathbf{f}_i$ is the $i^{th}$ row of the spectrum to be measured, $\mathbf{n}_{snap_i}$ is the measurement noise in $i^{th}$ row measurement. If the intensity at each element of the coded aperture is exactly the same, the normalized intensity $\mathbf{I}_S$ will turn into a standard Hadamard-S matrix, as described in $\mathbf{S} = \mathbf{I}_S / \max(\mathbf{I}_S) = \mathbf{S}_{snap}$. If the intensity of light is not uniform, the matrix will be approximate the standard Hadamard-S matrix, called sub-Hadamard-S matrix. Specifically, $\mathbf{S}_{snap}$ has a key feature that can be expressed as follows: $0 \leq \mathbf{S}_{snap}(i,j) \leq 1, \{i \leq n, j \leq n\}$. If each row of the scene has the same spectrum, or we want to measure the average spectrum of each column, i.e. $\mathbf{f}_1 = \mathbf{f}_2 = \cdots = \mathbf{f}_n$ or $\mathbf{f}_{11} = \mathbf{f}_{21} = \cdots = \mathbf{f}_{n1} = \frac{1}{n}\sum_{i=1}^{n}\mathbf{f}_{i1}$. Based on this assumption, the whole measurement of snapshot HTS can be simplified as:

$$\mathbf{g} = (\mathbf{S} \circ \mathbf{I})\mathbf{f} + \mathbf{n}_{snap} = \mathbf{S}_{snap}\mathbf{f} + \mathbf{n}_{snap} = (\mathbf{S} - \mathbf{S}_h)\mathbf{f} + \mathbf{n}_{snap} \qquad (2)$$

where, $\mathbf{S}_h$ is the difference between Hadamard-S matrix and normalized modulation intensity distribution, which satisfies $0 \leq \mathbf{S}_h(i,j) < 1$. Thus, once the light intensity distribution $\mathbf{S}_{snap}$ is obtained, the spectra need to measure can be recovered through an inverse process. The main difference between snapshot HTS and traditional multiple scan HTS is shown in the figure 2:

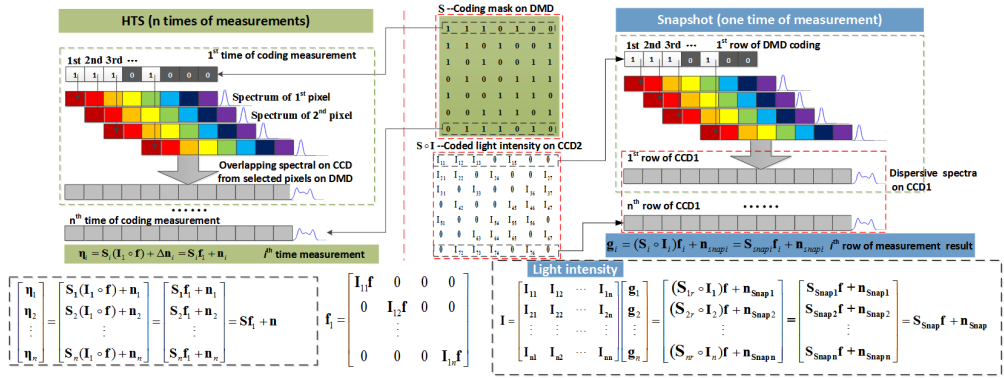

Figure 2. Principle difference between traditional HTS and sub-Hadamard snapshot spectrometer

In the traditional method, if we want to obtain the light intensity, we need an extra imaging path to capture the light intensity. However, as mentioned above, there are many problems in the dual-path scheme. Therefore, an appropriate method is needed to obtain the original intensity distribution from overlapped dispersive spectral images. For this reason, we propose a regression network to complete spectral unmixing.

## 3. Network building

As shown in Figure 1, in the raw spectral image, the intensity dispersion of each point is overlapped, and the captured spectral image can be expressed as the overlapped image of each band spectral image. This process can be expressed by the convolution operator. Each pixel in the spectrum can be represented as:

$$f_s = conv(p, S_{ij}) \tag{3}$$

where, $f_s$ is a single pixel in the spectral image, $p$ is a one-dimensional convolution kernel in the spectral direction, the kernel size of convolution is the number of spectral bands, and $S_{ij}$ is a high-dimensional spectral image.

From the perspective of spectral dimension, the light intensity of a point is the sum of its dispersive light intensity:

$$f_s = \sum_{i=1}^{n} S_i \tag{4}$$

Where $n$ is the single-pixel spectral dispersion length. This optical process is equal to the overlapping of a spectral image.

Therefore, we first design an unmixing module to reduce the mix of spectral image and then use an enhance module to recover the detail. The overall structure of the network is shown in Figure 3.

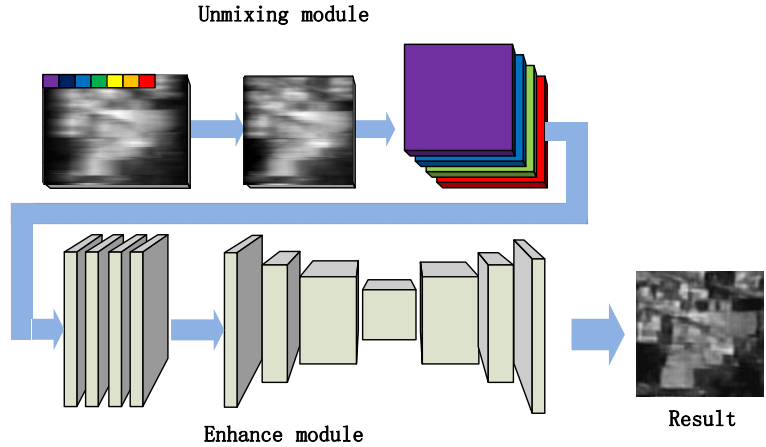

Figure 3. Structural sketch of network model

The unmixing module simulates the inverse process of superimposed and extracts the primary features through spectral direction convolution and deconvolution. In the enhance module, the output feature map is further enhanced to supplement more details such as image contrast, brightness, texture features and so on. The mean square error function is used as the loss function, and the Euclidean distance is expressed by $D_i = \|x_i - \hat{x}_i\|^2$. The total number of samples is N, then there are:

$$Loss(x_i, \hat{x}_i) = \frac{1}{N} \sum_{i=1}^{N} D_i \qquad (5)$$

where, $x_i, \hat{x}_i$ are real data and reconstructed data respectively. At the same time, regions with complex texture are harder to approximate, while the smooth region is easier to approximate. In order to make the training of networks focus on the harder samples, Wu Z et al put forward a method of hard-sample mining for pixel-level classification tasks [22]. Therefore, we adopt the idea and modify it to adapt our pixel-level regression task in training. We select pixel samples with higher loss values as hard negative samples for training and ignore those with lower loss values. So that the network will not repeatedly learn samples in low-frequency areas, and enhances the performance on high-frequency areas. In this case, the loss function can be expressed as:

$$Loss = \frac{1}{\sum_{i}^{N} 1\{D_i < t\}} \sum_{i}^{N} 1\{D_i < t\} D_i \qquad (6)$$

where, t is the threshold, which is the minimum value in the loss of the first half of the sample with larger loss value.

We use 1650 groups of multispectral images with 127×127 size for training, 350 groups for validation and 200 groups for testing. The images are padded to 128×128 for training conveniently. Finally, 128 ×(128+n) spectral dispersion map is constructed where *n* is the number of spectral bands. The images were superimposed by all bands are used as the training label.

In the unmixing module, the convolution layer with 1×n convolution kernel is used to unmixing, and the convolution layer with 3×3 convolution kernel and relu activation function are used to extract the primary features. In the enhancement module, the encoder-decoder network is used to extract the low-level and high-level semantic information of the unmixed feature map to enhance the unmixing image. The design of enhance module structure is referred to as ERFNet [23] shown in table 1. The downsampler layer is a pooling down-sampling process. The Non-bt-1d is the factorized convolution which decomposes the 3×3 convolution into a pair of 1D convolutions, which constitutes a non-bottleneck 1D structure as shown in Figure 4. With 33% reduction of parameters, we can achieve the same learning ability and accuracy as the traditional non-bottleneck structure, improve the network efficiency and contribute to real-time spectral analysis.

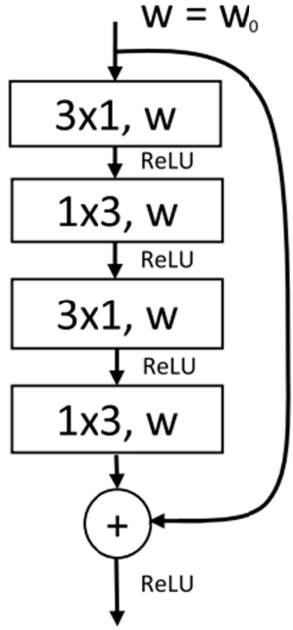

| Layer | Type | out-F | out-Res |
|---|---|---|---|
| 1 | conv(n*1) | 1 | 128*128 |
| 2 | deconv(1*1*n) | n | 128*128 |
| 3 | conv+relu | n/2 | 128*128 |
| 4 | conv+relu | n/4 | 128*128 |
| 5 | conv+relu | n/8 | 128*128 |
| 6 | conv+relu | 1 | 128*128 |
| 7 | Downsampler | 16 | 64*64 |
| 8 | Downsampler | 64 | 32*32 |
| 9--13 | 5*Non-bt-1D | 64 | 32*32 |
| 14 | Downsampler | 128 | 16*16 |
| 15 | Non-bt-1D | 128 | 16*16 |
| 16 | Non-bt-1D | 128 | 16*16 |
| 17 | Non-bt-1D | 128 | 16*16 |
| 18 | Non-bt-1D | 128 | 16*16 |
| 19 | Non-bt-1D | 128 | 16*16 |
| 20 | Non-bt-1D | 128 | 16*16 |
| 21 | Non-bt-1D | 128 | 16*16 |
| 22 | Non-bt-1D | 128 | 16*16 |
| 23 | deconv | 64 | 32*32 |
| 24--25 | 2*Non-bt-1D | 64 | 32*32 |
| 26 | deconv | 16 | 64*64 |
| 27--28 | 2*Non-bt-1D | 16 | 64*64 |
| 29 | deconv | C | 128*128 |

(Rows 1–6: SIMULATION; Rows 7–22: ENCODER; Rows 23–29: DECODER)

Figure 4. non-bottleneck 1D structure                Table 1. Network structure

### 4. SNR analysis

In order to analyze the noise reduction performance of the proposed method compared with the traditional dual-path sub-Hadamard snapshot spectral detection, we can set the sub-Hadamard matrix normalized intensity distribution as $S_{snap}$, $S_1$ is the difference between the matrix after network reconstruction and $S_{snap}$ ($S_1 \ll S_{snap}$). Then, the reconstructed spectra can be written as follows:

$$\hat{f} = f + (S_{snap} - S_1)^{-1} n_s \tag{7}$$

Based on Eq. (7), the SNR of reconstructed spectrum can be expressed as:

$$SNR_{\hat{f}} = 10 \log \left( \frac{f^T f}{n_s^T \left(S_{snap}^T - S_1^T\right)^{-1} \left(S_{snap} - S_1\right)^{-1} n_s} \bigg/ \frac{f^T f}{n_s^T n_s} \right) \tag{8}$$

Let $n_s' = \left(S_{snap} - S_1\right)^{-1} n_s$, the upper equation can be changed to:

$$SNR_{\hat{f}} = 10\log\left(\frac{n_s^T n_s}{n_s^T \left(S_{snap}^T - S_1^T\right)\left(S_{snap} - S_1\right)^{-1} n_s}\right)$$

$$= 10\log\left(\frac{\left(n_s'\right)^T \left(S_{snap}^T S_{snap} - S_{snap}^T S_1 - S_1^T S_{snap} + S_1^T S_1\right) n_s'}{\left(n_s'\right)^T n_s'}\right) \quad (9)$$

Suppose $S_{snap} - S_1$ as $\alpha S_{snap} + S_2$ ($0 < \alpha < 1$), so:

$$\begin{aligned} S_{snap} - S_1 &= \alpha S_{snap} + S_2 \\ (1-\alpha) S_{snap} &= S_1 + S_2 \\ k S_{snap} &= S_1 + S_2 \end{aligned} \quad (10)$$

where, $k$ denotes perturbations involving reconstruction errors and actual light intensity. It is known that $0 \le k < 1$. Therefore, Eq. (9) can be written as follows:

$$SNR_{\hat{f}} = 10\log\left(\frac{\left(n_s'\right)^T \left((1-k)^2 S_{snap}^T S_{snap} + ((1-k)/k) B\right) n_s'}{\left(n_s'\right)^T n_s'}\right) \quad (11)$$

$$B = \left(S_1^T S_2 + S_2^T S_1 + (2-k)/(1-k) S_2^T S_2\right)$$

Based on the previous work, the improvement of SNR is as follows:

$$SNR_{\hat{f}} \ge 10\log\left(\frac{\left(n_s'\right)^T \left((1-k)^2 S_{snap}^T S_{snap} + \frac{1-k}{k} k^2 S_{snap}^T S_{snap}\right) n_s'}{\left(n_s'\right)^T n_s'}\right)$$

$$= 10\log\left(\frac{\left(n_s'\right)^T S_{snap}^T S_{snap} n_s'}{\left(n_s'\right)^T n_s'}\right) + 10\log(1-k) \quad (12)$$

It can be seen that the reconstructed SNR of network decreases by about $10\log\left(\frac{1}{1-k}\right)$ (dB) compared with the sub-Hadamard matrix which can accurately obtain the intensity distribution. Compared with slit and snapshot HTS without light intensity information, it still has obvious advantages. At the same time, in practice, because the light intensity of the proposed method is doubled compared with that of the dual-path scheme, the actual spectrum can be expressed as:

$$\hat{f} = 2f + (S_{snap} - S_1)^{-1} n_s \quad (13)$$

After similar deduction, the final SNR can be expressed as:

$$SNR_{\hat{f}} \geq 10\log\left(\frac{(n_s')^T S_{snap}^T S_{snap} n_s'}{(n_s')^T n_s'}\right) + 10\log(1-k) + 10\log(2) \qquad (14)$$

Therefore, the proposed method has certain advantages over the dual optical path scheme.

## 5. Experiments and simulations

A variety of scenarios were used for training, and other scenario data that were not involved in the training are taken as the test data. By using two-step iterative soft threshold optimization (TwIST) to reconstruct Multi-frame CASSI [24] and reconstruction results used as a comparison. At the same time, the reconstruction quality is further improved by four-frame coding.

Considering the need for an original image for spectral reconstruction, the full-1 coded and Hadamard coded different coding matrices spectral images are reconstructed to compare the results. The spectral images before and after encoding are shown in Figure 5. The code uses Hadamard-S matrix.

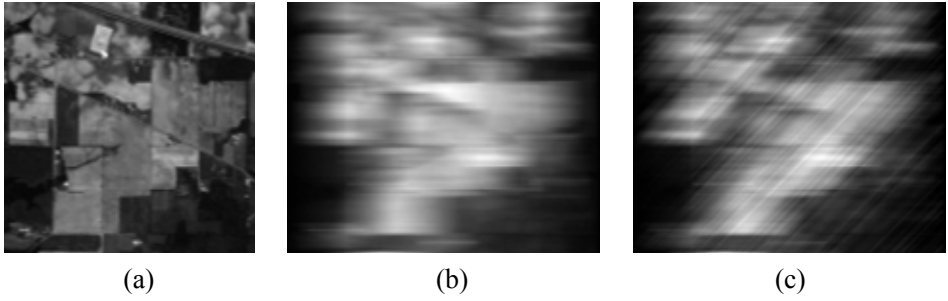

(a) (b) (c)

Figure 5. Hyperspectral data from lesun[25]. (a)The image of whole bands. (b) Full-1 code overlapping dispersion image. (c) Hadamard code overlapping dispersion image.

Different test data[25, 26] were reconstructed and compared peak signal-to-noise ratio (PSNR). The reconstruction results are shown in Figure 6.

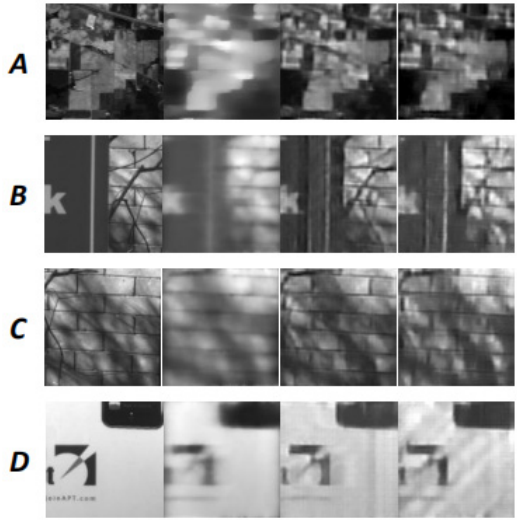

Figure 6. Reconstructed results. From left to right, they are: (1) original image; (2) reconstructed results of four frames CS; (3) reconstructed results of Full-1 coded spectrum; (4) reconstructed results of Hadamard coded spectrum;

| IMAGE | CASSI(4frame) | Full-1 | Hadamard |
|---|---|---|---|
| A | 12.3899 | 21.4796 | 21.1313 |
| B | 16.0883 | 22.3734 | 21.3922 |
| C | 19.4437 | 23.0336 | 22.901 |
| D | 20.508 | 20.815 | 20.6001 |

Table 2. Comparison of PSNR of Reconstruction of different scene.

It can be seen that the performance of the proposed method is stable under different conditions, and the reconstruction quality of the proposed method is better than that of the four frames CS in the case of a single frame. The test images are not involved in the training data, and the reconstructed online spectral database data can also get the expected results, which shows that the network has good generalization ability. At the same time, based on a desktop with Intel Core i7 8700K 4.3Ghz, 16GB memory and NVIDIA GeForce GTX 1060 6GB, the reconstructed time of GPU is 0.075s without additional hardware acceleration. If TensorFlow is used for acceleration and switch a higher performance GPU, the reconstructed speed will be significantly improved, which can basically meet the real-time requirements.

In order to verify the algorithm, we built a practical optical system. The spectrum with band-pass filters and the whole spectrum were trained and tested in the actual optical system to test the influence of overlapping degree on reconstruction effect. The optical system is shown in Figure 7, the incident light is encoded by a digital micromirror device (DMD) from

Texas Instrument (TI). The camera used in this paper is Basler acA1920-40um, and the dispersion grating uses THORLABS GT25-03(300 Grooves/mm,17.5deg). Among them, the DMD pixel size is 13.68um × 13.68um, and the camera pixel size is 5.86um × 5.86um.

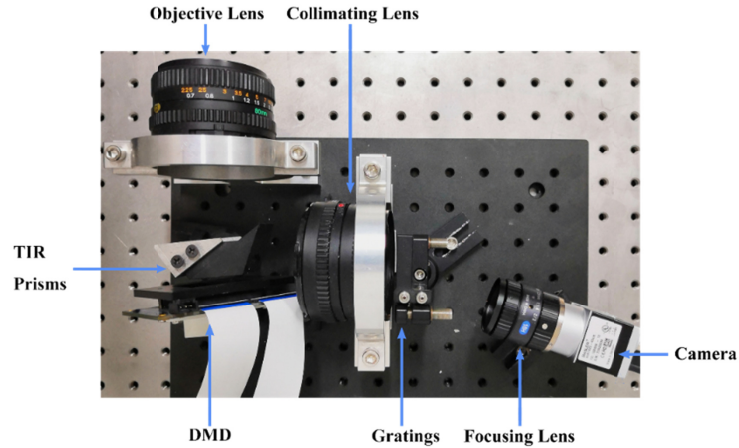

Figure 7. Optical system

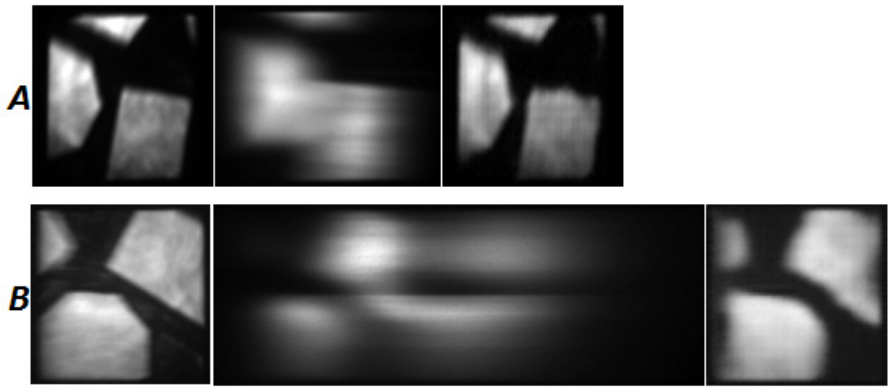

Figure 8. Reconstructed results of actual experiments. From left to right, these images are the original image, the dispersive image, and the reconstructed result. Among them, A uses a bandpass filter with a bandwidth of 100 nm (400-500nm) to truncate the spectrum, and B is a whole spectrum. PSNR (A) = 25.7557dB, PSNR (B) = 20.2598dB.

As shown in the Figure 8, it can be seen that in practice, our method can still reconstruct the original image well. However, with the increase of spectral bands, the overlapping becomes more serious, and the reconstruction effect will be affected to some extent.

In order to quantitatively analyze the effect of reconstructed light intensity on the reconstructed spectra, we compared the slit-based spectrometer, the snapshot HTS without light intensity, the sub-Hadamard snapshot spectrometer [1] and the proposed method. In the experiment, we ignore the light intensity difference in dual-path and single-path.

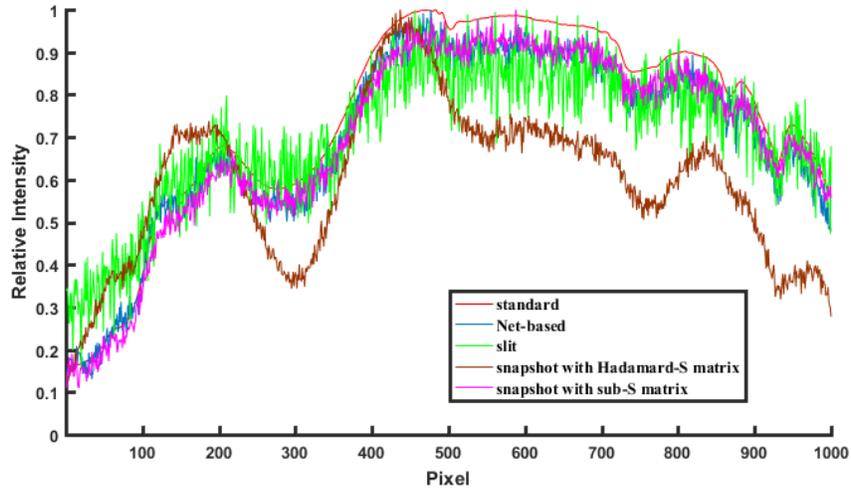

Figure 9. Spectral reconstruction simulation results. $SNR_{Net}$=15.0018dB, $SNR_{slit}$=8.4863dB, $SNR_{sHTS}$=5.5947dB, $SNR_{sub-S}$=16.1139dB.

From Figure 9, it can be found that the results of snapshot HTS without light intensity are seriously distorted due to the non-uniform intensity of light, which can be greatly improved by sub-Hadamard snapshot spectrometer. At the same time, the proposed method achieves similar results on the premise of reducing an optical path, while maintaining the advantages of slit spectrometer. It accords with the result of theoretical analysis.

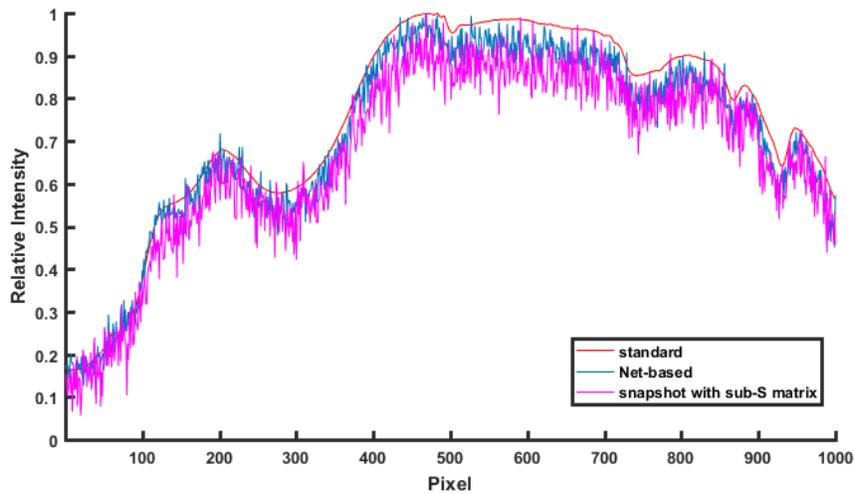

Figure 10. Comparison of the two-path scheme with the proposed method in the same case. $SNR_{Net}$=15.1712dB, $SNR_{sub-S}$=12.4685dB

At the same time, as shown in Figure 10, in practice, the dual-path sub-Hadamard snapshot spectrometer loses half of its light intensity, which will have a negative impact on the reconstruction of the SNR. According to the theoretical deduction, the SNR decline is:

$$SNR^- = 10\log(2) \tag{15}$$

Therefore, the performance of dual-path sub-Hadamard snapshot spectrometer is lower than the proposed method in practice.

In addition, we added the test of the influence of poor reconstruction image on the reconstruction effect. The original light intensity was reconstructed using the model which had not been well converged at the beginning of the network training. At this time, and the PSNR of the reconstructed light intensity decreased to 18.3746 dB, and the SNR of the reconstructed spectrum decrease to 14.0585 dB.

In view of the influence of the quality of network reconstruction on the spectral SNR, we made further comparisons, and the results are shown in Figure 11. Among them, the results of repeated reconstruction of the dual-path scheme are compared. It can be seen that as the number of iterations increases, the quality of light intensity reconstruction increases, and the SNR of reconstructed spectra also increases. As the quality of reconstruction increases, the advantage of light intensity gradually manifests, and the SNR is better than that of dual-path. It accords with the results of the previous theoretical derivation.

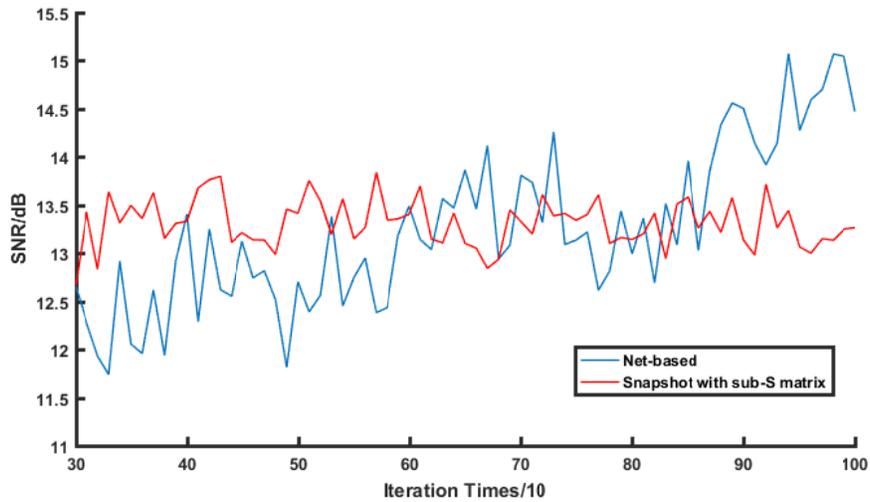

Figure 11. The relationship between the number of iterations of the network and the spectral SNR

In the actual experiment, we use the LED spectrum as the measurement target and use the gain of 30dB to increase the system noise and simulate the high noise environment. The captured spectral image is shown in Figure 12 for example. And the reconstructed spectral data are shown in Figure 13.

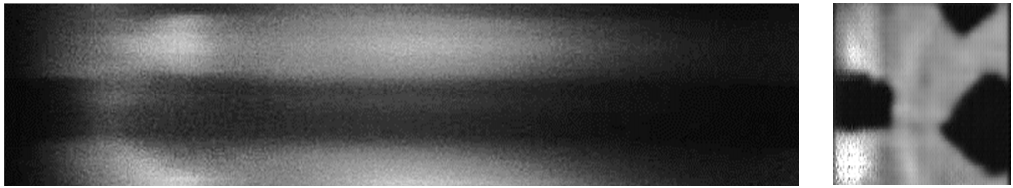

(a) (b)

Figure 12. The results of actual experiments, (a) Spectral images. (b) Reconstruction results

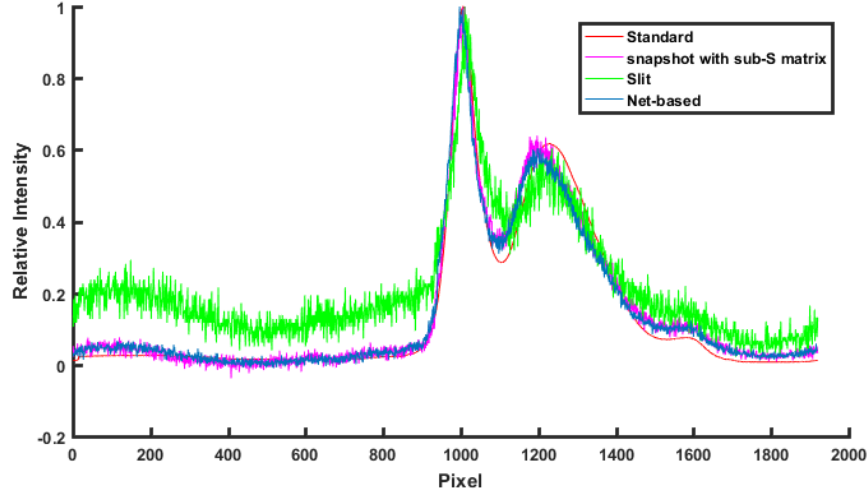

Figure 13. 30dB gain experimental results. $SNR_{Net}$=14.0912dB, $SNR_{slit}$=6.2915dB, $SNR_{sub-S}$=13.052dB

It can be seen that in the actual test, our scheme has a better SNR than the dual-path scheme because of the advantage of light throughput.

## 6. Conclusion

On the basis of sub-Hadamard snapshot spectral detection, this paper proposes an innovative method of spectral image unmixing using a convolutional neural network to obtain the original image and applies it to spectral detection. The proposed method can solve the drawbacks of sub-Hadamard spectrometer such as pixel-level registration, loss of light throughput, highly complex. The proposed method can obtain fairly or even better performance then dual-path sub-Hadamard snapshot spectrum detection under the premise of reducing one optical path, which further improves its practicability. At the same time, the proposed idea of spectral unmixing can be further developed in the field of Computational Spectral Imaging in the future.

**Acknowledgements:** This work was supported by the National Natural Science Foundation of China (Grant Nos. 61727802, 61601225). We thank Enlai Guo for technical supports.